\documentclass[11pt]{amsart}
\usepackage{amsxtra,amssymb,amsmath,amsthm,amsbsy}
\usepackage{latexsym}
\usepackage{rotating}
\usepackage{mathrsfs}
\usepackage{graphicx}

\usepackage[T1]{fontenc}
\usepackage[francais]{babel}

\begin{document}
\begin{titlepage}
\begin{center}
\Large
\textsc\textbf{MINISTERE DE L'EQUIPEMENT}\\
\textsc{Office de la Topographie et du Cadastre }\\

\vspace{5cm}

\LARGE
\textsc{\textbf{Interprétation Géométrique du Problème de Compensation Non-Linéaire}}\\
[0.5\baselineskip]
{Par \\ Abdelmajid BEN HADJ SALEM}\\
\vspace{0.5cm}
\normalsize
\textsc{Ingénieur Général à l'Office de la Topographie et du Cadastre}\\

\vspace{1cm}
\textsc{Septembre 2011}
\\ 

\vspace{1cm}
\textsc{Version 1.}\\

\vspace{5cm}
\textsc{Office de la Topographie et du Cadastre}\\
\textsl{www.otc.nat.tn}

\end{center}
\end{titlepage}
\tableofcontents
\clearpage 
\Large
\begin{center}
\textbf{Interprétation Géométrique du Problème de Compensation Non-Linéaire}
\\
\vspace{2mm}

\textsc{\textbf{Abdelmajid Ben Hadj Salem}}

\end{center}
\date{}


\normalsize
\vspace{6mm}
\section{Introduction}

Dans un article \cite{bib1} E. Grafarend et B. Schaffrin ont étudié la géométrie de la compensation ou l'ajustement non-linéaire et ont présenté le cas du problème d'intersection plane en utilisant le modèle de Gauss Markov, par les moindres carrés. Le présent papier présente les principes de la géométrie de la  compensation non-linéaire par la méthode des moindres carrés en s'appuyant sur le Lemme de P\'azman \cite{bib2}.

\vspace{4mm}
\section{La Géométrie Non Linéaire du Modèle de Gauss-Markov}
Le modèle non linéaire de Gauss-Markov est défini par:
\begin{equation}
	\zeta(X)=L-e; \quad e \in \mathcal{N}(0,\Gamma) \label{m1}
\end{equation}
avec:

- $L$: le vecteur des observations $(n\times1)=(L_1,L_2,..,L_n)^T$,

- $X$: le vecteur des inconnues $(m\times1)=(X_1,X_2,..,X_m)^T$,

- $e$: le vecteur des erreurs $(n\times1)=(e_1,e_2,..,e_n)^T$ suit la loi normale $\mathcal{N}(0,\Gamma)$ avec $E(e)=0$ et $\Gamma=E(ee^T)$ la matrice de dispersion ou variance, on prendra $\Gamma=\sigma^2_0.P^{-1}$. $P$ est la matrice des poids et $\sigma_0$ une constante positive.

- $\zeta$: est une fonction donnée injective d'un ouvert $U\subset \mathbb R^m \rightarrow \mathbb R^n$ et  $m<n$.
\\

Remarque: dans le cas d'un modèle linéaire, la fonction $\zeta=A.X$ où $A$ est une matrice $n\times m$.
\\

On note $Im \zeta=\left\{\zeta(X) / X \in U\right\}$ l'image de $U$ par la fonction $\zeta$. $Im\zeta$ est une variété de dimension $m$ vérifiant les conditions:
\vspace{4mm}

(i): les vecteurs $\displaystyle \frac{\partial \zeta}{\partial X_1},\frac{\partial \zeta}{\partial X_2},...,\frac{\partial \zeta}{\partial X_m}$ sont linéairement indépendants en chaque point $X \in U$,
\vspace{4mm}

(ii): les fonctions $\displaystyle \frac{\partial^2 \zeta}{\partial X_i \partial X_j}$ sont continues sur $U$ pour $i,j \in \left\{1,2,...,m\right\}$.
\\

On introduit un produit scalaire:
\begin{equation}
	<\zeta_1,\zeta_2>=\zeta_1^T.P.\zeta_2 \label{m2}
\end{equation}
 D'où la norme du vecteur $\zeta=(\zeta_1,\zeta_2,...,\zeta_n)^T$:
 \begin{equation}
	\|\zeta \|^2=<\zeta,\zeta>=\zeta^T.P.\zeta =\sum_{i=1}^np_i.\zeta_i^2 \label{m3}
\end{equation}
dans l'espace vectoriel $\mathbb R^n$ en prenant la matrice de poids $P$ une matrice diagonale . 
\\

Alors la solution par les moindres carrés $\bar{X}$ sera définie par:
\begin{equation}
		\|L-\bar{\zeta}(\bar{X}) \|=min\left\{\|L-\zeta(X) \| \,/ \,X\in U \right\} \label{m4}
\end{equation}
Cette condition est exprimée par les équations de Lagrange-Euler soit:
\begin{equation}
	\frac{\partial}{\partial X_i} \|L-\zeta(X) \|^2=0 \quad \hbox{pour}\,i\in \left\{1,2,...,m\right\} \label{m5}
\end{equation}
En effet, on veut minimiser la fonction:
\begin{equation}
	F(X)=F(X_1,X_2,..,X_m)=\|L-\zeta(X) \|=\|L-\zeta(X_1,X_2,...,X_m) \| \label{m6}
\end{equation}
Comme $F$ est une fonction positive, minimiser $F$ c'est aussi minimiser $F^2$, soit $J(X)=F^2(X)$. En appliquant les équations de Lagrange-Euler, on obtient:
$$	-\frac{\partial J(X)}{\partial X_i}=0 \Rightarrow \frac{\partial J(X)}{\partial X_i}=0 $$ 
soit:
\begin{equation}
	\frac{\partial}{\partial X_i} \|L-\zeta(X_1,X_2,...,X_m) \|^2=0 \quad \hbox{pour}\,i\in \left\{1,2,...,m\right\} \label{m7}
\end{equation}
or:
\begin{align}
 \|L-\zeta(X_1,X_2,...,X_m) \|^2=(L-\zeta(X_1,X_2,...,X_m))^T.P.(L-\zeta(X_1,X_2,...,X_m))= 	\nonumber \\
	 \zeta(X)^T.P.\zeta(X) -2L^T.P.\zeta(X)+L^T.P.L \label{m8} 
	\end{align}
Soit :
\begin{equation}
		\frac{\partial  J(X)}{\partial X_i}=2\zeta(X)^T.P.	\frac{\partial \zeta(X)}{\partial X_i}-2L^T.P.\frac{\partial \zeta(X)}{\partial X_i} \quad \hbox{pour}\,i\in \left\{1,2,...,m\right\} \label{m9}
\end{equation}
ou encore :
$$ 		\frac{\partial  J(X)}{\partial X_i}=2(\zeta(X)-L)^T.P.\frac{\partial \zeta(X)}{\partial X_i} \quad \hbox{pour}\,i\in \left\{1,2,...,m\right\} $$

ce qui donne en utilisant \eqref{m5}:
\begin{align}
& \label{m10} <L-\zeta(X),\frac{\partial \zeta(X)}{\partial X_i}>=0 \quad \hbox{pour}\,i\in \left\{1,2,...,m\right\}  \\
& \label{m11} \hbox{ou}\quad <e,\frac{\partial \zeta(X)}{\partial X_i}>=0 \quad \hbox{pour}\,i\in \left\{1,2,...,m\right\}
 \end{align}
Géométriquement, cela veut dire que le vecteur erreur $e =L-\zeta(X)$ est perpendiculaire (produit scalaire nul) au plan tangent de la variété $Im \zeta$ au point $\bar{\zeta}(\bar{X})$ (s'il existe).
\\

Pour le cas non-linéaire, la condition \eqref{m10} est nécessaire mais non suffisante. Pour obtenir le minimum, il faut que la matrice $(\displaystyle \frac{\partial^2 J}{\partial X_i \partial X_j} ),i,j\in \left\{1,2,...,m\right\} $ soit définie positive.

\section{Interprétation Géométrique}
Dans cette section, on va clarifier l'interprétation géométrique de façon que la solution de \eqref{m10} soit localement unique.
\\

Considérons la matrice $m\times m$ définie par:
\begin{equation}
	G(X)=(G_{\alpha\,\beta}) \quad \hbox{avec}\,G_{\alpha\,\beta}=<\frac{\partial \zeta(X)}{\partial X_{\alpha}},\frac{\partial \zeta(X)}{\partial X_{\beta}}> \quad 
\begin{pmatrix}
	\alpha=1,2,..,m \\
	\beta=1,2,...,m
\end{pmatrix} \label{m12}
\end{equation}
Or:
\begin{equation}
	 ds^2=G_{\alpha\,\beta}dX_{\alpha}dX_{\beta} \label{m13}
	 \end{equation}
représente la métrique de la variété $Im \zeta$. La matrice $G(X)= (G_{\alpha\,\beta})$ est appelée en terme statistique la matrice d'information de Fischer. 
\\

Introduisons la matrice $B$ définie par:
\begin{equation}
	B(X,L)=(B_{\alpha\,\beta}) \quad \hbox{avec}\,B_{\alpha\,\beta}=\frac{1}{2}\frac{\partial^2 }{\partial X_{\alpha}\partial X_{\beta}}\|L-\zeta(X)\|^2 \quad 
\begin{pmatrix}
	\alpha=1,2,..,m \\
	\beta=1,2,...,m
\end{pmatrix} \label{m14}
\end{equation}
 or: 
\begin{align}
	 \|L-\zeta(X)\|^2=(L-\zeta(X))^TP(L-\zeta(X))=(L^T-\zeta^T(X))(PL-P\zeta(X))=\nonumber \\ L^TPL-2L^TP\zeta(X)+\zeta^T(X)P\zeta(X)\label{m15}
\end{align}
 D'où:
 \begin{align}
	\frac{\partial}{\partial X_{\alpha}}(L^TPL-2L^TP\zeta(X)+\zeta^T(X)P\zeta(X))=-2L^TP\frac{\partial \zeta}{\partial X_{\alpha}}+2\zeta^T(X)P\frac{\partial \zeta(X)}{\partial X_{\alpha}}= \nonumber \\ -2(L^T-\zeta^T(X))P\frac{\partial \zeta(X)}{\partial X_{\alpha}} \label{m16}
\end{align}
Donc:
\begin{align}
	\frac{1}{2}\frac{\partial^2 }{\partial X_{\alpha}\partial X_{\beta}}\|L-\zeta(X)\|^2=-L^TP\frac{\partial ^2 \zeta(X)}{\partial X_{\alpha}\partial X_{\beta} }+\frac{\partial \zeta(X)}{\partial X_{\beta}}P\frac{\partial \zeta(X)}{\partial X_{\alpha}}+\zeta^T(X)P\frac{\partial^2 \zeta(X)}{\partial X_{\alpha} \partial X_{\beta}}= \nonumber \\ \frac{\partial \zeta^T(X)}{\partial X_{\beta}}P\frac{\partial \zeta(X)}{\partial X_{\alpha}}-(L^T-\zeta^T(X))P\frac{\partial^2 \zeta(X)}{\partial X_{\alpha} \partial X_{\beta}}= \nonumber \\ <\frac{\partial \zeta(X)}{\partial X_{\beta}},\frac{\partial \zeta(X)}{\partial X_{\alpha}}> -<L-\zeta(X),\frac{\partial ^2 \zeta(X)}{\partial X_{\beta}\partial X_{\alpha}}> \label{m17}
\end{align}
Soit:
\begin{equation}
	B_{\alpha\,\beta}=G_{\alpha\,\beta}-<L-\zeta(X),\frac{\partial ^2 \zeta(X)}{\partial X_{\beta}\partial X_{\alpha}}> \label{m18}
\end{equation}
Posons:
\begin{equation}
	H(L,X)=(h_{\alpha \beta})=(<L-\zeta(X),\frac{\partial ^2 \zeta(X)}{\partial X_{\beta}\partial X_{\alpha}}>)\quad 
\begin{pmatrix}
	\alpha=1,2,..,m \\
	\beta=1,2,...,m
\end{pmatrix} \label{m19} 
\end{equation}
c'est-à-dire:
\begin{equation}
	B=G-H \label{m20}
\end{equation}
Revenons à $Im\zeta(X)=\left\{ \zeta(X)\quad / X\in U\right\}$. Soit une ligne géodésique de $Im \zeta(X)$ passant par un point $\zeta=\zeta(X)$ paramétrée par son abcisse curviligne $s$, on a alors:
\begin{equation}\label{m21}
  \chi(s)=\zeta(X(s)),\,\,\,s\in [s_1,s_2] 
\end{equation}
où $X(s)$ décrit une certaine courbe dans le domaine $U\subset \mathbb R^m$.
\\

Le vecteur :
\begin{equation}
	\chi'(s)=\frac{ d\chi(s)}{ds} \label{m22}
\end{equation}
représente le vecteur tangent à la ligne géodésique au point $\zeta(X(s))$ de $Im \zeta(X)$. Ce vecteur vérifie:
\begin{equation}
	\|\chi'(s)\|^2=1  \label{m23}
\end{equation}
Par suite, la dérivée de ce vecteur par rapport à $s$ est un vecteur orthogonal à $\chi'(s)$ donc orthogonal à $Im \zeta(X(s))$ au point $\zeta(X(s))$:
\begin{equation}
		\chi"(s)=\frac{ d\chi'(s)}{ds} \bot \,\chi'(s)  \label{m24}
\end{equation}
 c'est-à-dire parallèle au vecteur normal à la surface ou la variété $Im \zeta(X)$ et on retrouve la propriété que $\chi(s)$ est une géodésique.
 
 Remarquons que pour une ligne géodésique, la courbure géodésique est nulle et la courbure normale coincide avec la courbure de la courbe $\chi(s)$ soit:
\begin{equation}
	\rho(s)=\frac{1}{\|\chi"(s)\|} \label{m25}
\end{equation}
le rayon de courbure. Appelons:
\begin{equation}
	n(s)=\frac{\chi"(s)}{\|\chi"(s)\|}=\chi"(s).\rho(s) \label{m26}
\end{equation}
c'est un vecteur unitaire perpendiculaire au plan tangent à la surface $Im \zeta(X)$.

D'après l'équation \eqref{m10}, au point $\zeta(\bar{X})$, le vecteur $e$ est perpendiculaire à $L-\zeta(\bar{X})$. Appelons alors : 
  \begin{equation}
	K(\zeta(\bar{X}))=\left\{ Z /\,Z\in \mathbb R^n \hbox{avec} <Z,\frac{\partial \zeta(\bar{X})}{\partial X_{\alpha}}> =0\,\,\alpha=1,2,...,m \right\} \label{m27} 
\end{equation}
On donc $e=L-\zeta(\bar{X}) \in K$. Ce dernier est un espace vectoriel de dimension $n-m$ orthogonal à $Im\zeta(X)$ au point $\zeta(\bar{X})$. On a aussi $\chi"(s) \in K$.
\subsection{Lemme de P\'azman}
 On peut maintenant énoncer le lemme de P\'azman (1984,\cite{bib2}) comme suit:
 
 \textbf{Lemme de P\'azman:}
  \textit{ Pour tout vecteur d'observation $L \in \mathbb R^m $, et toute solution appropriée $\bar{X}$ des équations:
  $$  <L-\zeta(\bar{X}),\frac{\partial \zeta(\bar{X})}{\partial X_{\alpha}}> =0,\,\,\alpha=1,2,...,m $$}
  
\textit{les conditions suivantes sont équivalentes:}
  
 \textit{1)- La matrice} $B(\bar{X},L)=G(\bar{X})-(<L-\zeta(\bar{X}),\frac{\partial ^2\zeta(\bar{X})}{\partial X_{\alpha} \partial X_{\beta}}>)$ \textit{est définie positive.}
 
\textit{ 2)- Pour toute ligne géodésique $\chi(X(s))$ vérifiant :}
 $$ \chi(\bar{s})=\zeta(\bar{X}(\bar{s})) $$
\textit{On a l'inégalité:}
\begin{equation}
	 <L-\zeta(\bar{X}),n(\bar{s})>\,\, < \, \rho(\bar{s}) \label{m28}
\end{equation}
\\

\textbf{A. La condition 1:}

En effet, supposons que la matrice $B(\bar{X},L)$ est définie positive c'est-à-dire:
 \begin{equation}\label{m29}
	\forall Y \in \mathbb R^m \,\, , Y\neq 0 \Rightarrow Y^T.B.Y > 0 
	\end{equation}
	Prenons alors :$Y=\chi'(\bar{s}) $. On a:
	\begin{equation}
	\chi'^T(\bar{s}).B(L,\bar{X}).\chi'(\bar{s}) > \,0  \label{m30}
\end{equation}
Comme $B=G(\bar{X}) - H(L,\bar{X})$, on obtient:
$$\chi'^T(\bar{s}).(G(\bar{X})- H(L,\bar{X})).\chi'(\bar{s}) > \,0 $$
soit:
\begin{equation}
	\chi'^T(\bar{s}).G(\bar{X})\chi'(\bar{s})- \chi'^T(\bar{s})H(L,\bar{X}).\chi'(\bar{s}) > \,0 \label{m31}
\end{equation}
Or pour $s$:
\begin{equation}
 Y=\chi'(s)=\frac{d\chi(s)}{ds}=\sum_{i=1}^{i=m}\frac{\partial \zeta(X(s))}{\partial X_i}\frac{d X_i(s)}{ds}=\sum_{i=1}^{i=m}\chi'_i(s)\frac{\partial \zeta(X(s))}{\partial X_i} \label{m32}
\end{equation}
en notant $\chi'_i(s)=\frac{d X_i(s)}{ds}$ les composantes de $\chi'(s)$ dans le plan tangent à $Im \zeta$ au point $\zeta (X(s))$. Comme:
\begin{align}
\| \chi'(\bar{s})\|^2=1=\chi'^T(s).\chi'(s)= <\sum_{i=1}^{i=m}\chi'_i(s)\frac{\partial \zeta(X(s))}{\partial X_i},\sum_{j=1}^{j=m}\chi'_i(s)\frac{\partial \zeta(X(s))}{\partial X_i}>=\nonumber \\ \sum_{i,j=1}^{m}\chi'_i(s).\left( <\frac{\partial \zeta(X(s))}{\partial X_i},\frac{\partial \zeta(X(s))}{\partial X_j}> \right)\chi'_j(s) =\chi'(s)^T.G.\chi'(s)=1 \label{m33}
\end{align}
Prenons $s=\bar{s}$, alors \eqref{m31} devient:
\begin{equation}
	 \chi'^T(\bar{s})H(L,\bar{X}).\chi'(\bar{s}) < \,1 \label{m34}
\end{equation}
Comme $B=G-H$  donc la matrice $H$ est exprimée dans la base de $B$ soit $\left(\frac{\partial \zeta(X(s))}{\partial X_j}\right)$
En utilisant \eqref{m32}, le nombre réel $\chi'^T(\bar{s})H(L,\bar{X}).\chi'(\bar{s})$ s'écrit:
\begin{equation}
\chi'^T(\bar{s})H(L,\bar{X}).\chi'(\bar{s})=\sum_{i=1}^{i=m}\chi'_i(\bar{s})\left(\sum_{j=1}^{j=m}h_{ij}.\chi'_j(\bar{s})\right)=\sum_{i,j=1}^{m}\chi'_i(\bar{s}).\chi'_j(\bar{s}).h_{ij} \label{m35}
\end{equation}
On remplace $h_{ij}$ par $$<L-\zeta(\bar{X}),\frac{\partial ^2\zeta(\bar{X})}{\partial X_i \partial X_j}> $$ Par un calcul simple, l'équation \eqref{m35} devient:
\begin{equation}
\chi'^T(\bar{s})H(L,\bar{X}).\chi'(\bar{s})= <L-\zeta(\bar{X}),\sum_{i,j=1}^{m}\chi'_i(\bar{s}).\chi'_j(\bar{s})\frac{\partial ^2\zeta(\bar{X})}{\partial X_i \partial X_j}> \label{m36}
\end{equation}
Maintenant, on va s'intéresser au membre droit du produit scalaire de l'équation \eqref{m36}. En différentiant l'équation \eqref{m32} par rapport à $s$, on obtient:
\begin{equation}
	\chi"(s)=\sum_i \frac{d\chi'_i(s)}{ds}.\frac{\partial \zeta(X(s))}{\partial X_i}+\sum_i\chi'_i(s)\sum_j\frac{\partial^2 \zeta(X(s))}{\partial X_i \partial X_j}.\frac{dX_j}{ds} \label{m37}
	\end{equation}

	Alors on a pour $s=\bar{s}$:
	\begin{align}
<L-\zeta(\bar{X}),\chi"(\bar{s})>=<L-\zeta(\bar{X}),\sum_i \frac{d\chi'_i(s)}{ds}.\frac{\partial \zeta(X(s))}{\partial X_i}+\sum_i \chi'_i(s)\sum_j \frac{\partial^2 \zeta(X(s))}{\partial X_i \partial X_j}.\frac{dX_j}{ds}>= \nonumber \\
\sum_i \frac{d\chi'_i(s)}{ds}.<L-\zeta(\bar{X}),\frac{\partial \zeta(X(s))}{\partial X_i}>+\sum_i \chi'_i(s)\sum_j <L-\zeta(\bar{X}),\frac{\partial^2 \zeta(X(s))}{\partial X_i \partial X_j}.\frac{dX_j}{ds}> \label{m38}
\end{align}
Or en utilisant l'équation \eqref{m10}, le premier terme de la deuxième ligne de l'équation précédente est nul:
$$ \sum_i \frac{d\chi'_i(s)}{ds}.<L-\zeta(\bar{X}),\frac{\partial \zeta(X(\bar{s}))}{\partial X_i}> =0 $$
 et comme :
$$ \chi'_j(s)=\frac{dX_j}{ds}$$
Alors l'équation \eqref{m38} devient:
	\begin{align}
<L-\zeta(\bar{X}),\chi"(\bar{s})>=\sum_i \chi'_i(s)\sum_j <L-\zeta(\bar{X}),\frac{\partial^2 \zeta(X(s))}{\partial X_i \partial X_j}.\chi'_j(s)> = \nonumber \\ \sum_i \sum_j \chi'_j(\bar{s})\chi'_i(\bar{s}) <L-\zeta(\bar{X}),\frac{\partial^2 \zeta(X(\bar{s}))}{\partial X_i \partial X_j}> \label{m39}
\end{align}
Or le deuxième membre n'est autre que l'équation \eqref{m36}. En utilisant \eqref{m34}, on obtient:
\begin{equation}
	<L-\zeta(\bar{X}),\chi"(\bar{s})> \,\,< 1 \label{m40}
\end{equation}
Or:
$$ n(\bar{s})=\rho(\bar{s}).\chi"(\bar{s})$$
D'où:
\begin{equation}
	<L-\zeta(\bar{X}),n(\bar{s})> \quad < \, \rho(\bar{s}) \label{m41}
\end{equation}

\textbf{B. La condition 2:}

  On suppose que reciproquement, on a pour toute géodésique $ \chi (s)=\zeta(X(s))$ de $ Im \zeta $ passant par le point $ \chi(\bar{s})=\zeta(\bar{X}(\bar{s}))$ vérifiant:
\begin{equation}
	<L-\zeta(\bar{X}),n(\bar{s})>\quad< \, \rho(\bar{s}) \label{m42}
\end{equation}
et :
$$ <L-\zeta(\bar{X}),\frac{\partial \zeta(X)}{\partial X_i}>=0 \quad \hbox{pour}\,i\in \left\{1,2,...,m\right\} $$
telque $\chi'(s)$ vérifiant :$$ \|\chi'(\bar{s}) \|^2=1 $$
Comme: $$ n(s)=\frac{\chi"(s)}{\|\chi"(s)\|}=\rho(s)\chi"(s)$$ 
le remplaçant dans l'équation \eqref{m42}, on obtient:
\begin{equation}
	<L-\zeta(\bar{X}),\rho(s)\chi"(s)>\quad <\, \rho(\bar{s}) \label{m43}
\end{equation}
et en simplifiant par $\rho \neq 0$, soit:
\begin{equation}
	<L-\zeta(\bar{X}),\chi"(s)>\quad < \,1 \label{m44}
\end{equation}
Comme:
\begin{eqnarray*}
	 \chi(s)=\zeta(X(s))\Rightarrow \chi'(s)=\frac{d\chi(s)}{ds}=\sum_{i=1,m}\frac{\partial \zeta}{\partial X_i}\frac{dX_i(s)}{ds} \\
	=\sum_{i=1,m}\chi'_i(s)\frac{\partial \zeta}{\partial X_i} 
\end{eqnarray*}
D'où en dérivant une deuxième fois par rapport à $s$:
\begin{align}
		\chi"(s)=\sum_i \frac{d\chi'_i(s)}{ds}.\frac{\partial \zeta(X(s))}{\partial X_i}+\sum_i\chi'_i(s)\sum_j\frac{\partial^2 \zeta(X(s))}{\partial X_i \partial X_j}.\frac{dX_j}{ds} \nonumber \\
	 = \sum_i \frac{d\chi'_i(s)}{ds}.\frac{\partial \zeta(X(s))}{\partial X_i}+\sum_i\sum_j \chi'_i(s)\chi'_j(s)\frac{\partial^2 \zeta(X(s))}{\partial X_i \partial X_j} \label{m45}
\end{align}
En remplaçant $\chi"(s)$ dans \eqref{m44}, on obtient:
\begin{equation}
	<L-\zeta(\bar{X}),\sum_i \frac{d\chi'_i(s)}{ds}.\frac{\partial \zeta(X(s))}{\partial X_i}+\sum_i\sum_j \chi'_i(s)\chi'_j(s)\frac{\partial^2 \zeta(X(s))}{\partial X_i \partial X_j}>\quad < \,1 \label{m46}
\end{equation}
ou encore:
\begin{equation}
		\sum_i\chi"_i(s)<L-\zeta(\bar{X}),\frac{\partial \zeta(X(s))}{\partial X_i}>+\sum_i\sum_j \chi'_i(s)\chi'_j(s)<L-\zeta(\bar{X}),\frac{\partial^2 \zeta(X(s))}{\partial X_i \partial X_j}>\quad < \,1 \label{m47}
\end{equation}
Or la première somme est nulle en vertu de l'équation \eqref{m10}. Il reste:
\begin{equation}
	\sum_i\sum_j \chi'_i(s)\chi'_j(s)<L-\zeta(\bar{X}),\frac{\partial^2 \zeta(X(s))}{\partial X_i \partial X_j}>\quad < \,1 \label{m48}
\end{equation}
Comme:
\begin{eqnarray*}
	 \chi'(s)=\frac{d\chi(s)}{ds}=\sum_{i=1,m}\frac{\partial \zeta}{\partial X_i}\frac{dX_i(s)}{ds} \\
	=\sum_{i=1,m}\chi'_i(s)\frac{\partial \zeta}{\partial X_i} 
\end{eqnarray*}
et $\chi'(s)$ vérifie $\|\chi'(s)\|=1$ car c'est un vecteur unitaire tangent à la géodésique $\chi(s)$. Donc
\begin{equation}
		<\chi'(s),\chi'(s)>=1 \Rightarrow	<\sum_{i=1,m}\chi'_i(s)\frac{\partial \zeta}{\partial X_i},\sum_{j=1,m}\chi'_j(s)\frac{\partial \zeta}{\partial X_j}>=1 \label{m49}
\end{equation}
soit:
\begin{equation}
	\sum_{i=1,m}\sum_{j=1,m}\chi'_i(s)	<\frac{\partial \zeta}{\partial X_i},\frac{\partial \zeta}{\partial X_j}>\chi'_j(s)=	\sum_{i=1,m}\sum_{j=1,m}\chi'_i(s)G_{ij}\chi'_j(s)=1 \label{m50}
\end{equation}
et ce-ci n'est autre que :
\begin{equation}
	\chi'^T(s).G.\chi'(s)=1 \label{m51}
\end{equation}
En utilisant l'équation \eqref{m48}, on a:
\begin{equation}
	\sum_i\sum_j \chi'_i(s)\chi'_j(s)<L-\zeta(\bar{X}),\frac{\partial^2 \zeta(X(s))}{\partial X_i \partial X_j}>\quad < \,\chi'^T(s).G.\chi'(s) \label{m52}
\end{equation}
ou encore:
\begin{equation}
0< \chi'^T(s).G.\chi'(s)-	\sum_i\sum_j \chi'_i(s).h_{ij}.\chi'_j(s) \label{m53}
\end{equation}
Finalement, nous obtenons:
\begin{equation}
0< \chi'^T(s).G.\chi'(s)-	\chi'^T(s).H.\chi'(s) \label{m54}
\end{equation}
C'est-à-dire pour tout vecteur $Y=\chi'(s)\neq 0$ du plan tangent de $Im \zeta$:
\begin{equation}
	Y^T.B.Y > \quad 0 \Longrightarrow \quad \hbox{ la matrice $B$ est définie positive} \label{m55}
\end{equation}
\begin{flushright}
C.Q.F.D
\end{flushright}
Maintenant, on peut dire quand $\bar{X}$ solution de \eqref{m10} est solution des moindres carrés en référence à l'équation \eqref{m4} et ce à partir du corollaire suivant:
\\

\textbf{Corollaire:}\textit{ Si $\bar{X}$ est solution de:}
\begin{equation}
<L-\zeta(X),\frac{\partial \zeta(X)}{\partial X_i}>=0 \quad \hbox{pour}\,i\in \left\{1,2,...,m\right\} \label{m56} 
\end{equation} 
\textit{ avec :}
\begin{equation}
	\|L-\zeta(\bar{X}\| < r \label{m57}
\end{equation}
\textit{où $r$ désigne le rayon de courbure minimum de la  variété $Im \zeta$ défini par:}
\begin{equation}
	r=inf\left\{ \rho_{\chi}(s)/\chi(s)=\zeta(X(s)) \,\,\hbox{une géodésique passant par}\,\zeta(X),\forall X\in U \right\} \label{m58}
\end{equation}
\textit{Alors $\bar{X}$ coincide avec la solution des moindres carrés $\bar{X}=\hat{X}(L)$.}
\\

\textbf{Démonstration:}

Comme les deux vecteurs $e=L-\zeta(\bar{X})$ et $n(\bar{s})$ sont orthogonaux à $Im \zeta$, ils sont colinéaires et comme $n(\bar{s})$ est un vecteur unitaire alors leur produit scalaire $<L-\zeta(\bar{X}),n(\bar{s})>$  est plus petit ou égal à $\|L-\zeta(\bar{X}\|$. Or ce terme est plus petit que $r$ d'après \eqref{m57}. et comme $r$ est le plus petit rayon de courbure, on a $r \leq \rho_{\chi}(\bar{s})$  pour toute géodésique $\chi(\bar{s})$ passant par $\zeta(\bar{X})$. Ce-ci est traduit par l'équation:
\begin{equation}
	<L-\zeta(\bar{X}),n(\bar{s})>\leq\|L-\zeta(\bar{X}\| < r \leq \rho_{\chi}(\bar{s}) \label{m59}
\end{equation}
 De cette dernière équation, on a:
 \begin{equation}
	<L-\zeta(\bar{X}),n(\bar{s})> \leq \rho_{\chi}(\bar{s}) \label{m50}
\end{equation}
En utilisant le lemme de P\'azman cité ci-dessus, la matrice $B$ est définie positive donc $\zeta(\bar{X})$ est un minimum strict \cite{bib3}. Or on a supposé que l'application $\zeta$ est injective (si $\zeta(X_1)=\zeta(X_2)\Longrightarrow X_1=X_2$), alors $\bar{X}$ coincide avec la solution des moindres carrés $\bar{X}=\hat{X}(L)$.

\addtocontents{toc}{\protect\vspace{\baselineskip}}


\begin{thebibliography}{99}

\bibitem{bib1} \textbf{E.W. Grafarend et B. Schaffrin}. The geometry of non-linear adjustment - the planar trisection problem. \textit{FESTCHRIFT to TORBEN KRARUP} edited by E. Kejlo, K. Poder and C.C. Tscherning. Geod\ae tisk Institut, Meddelelse n° 58. p 149-172. K\o benhavn, Danmark. 1989. 

\bibitem{bib2} \textbf{A. P\'azman, 1984}. Probability distribution of the multivariate nonlinear least-squares estimates; Kybernetika 20 (1984), p 209-230.

\bibitem{bib3} \textbf{ H. Cartan}. Cours de Calcul Différentiel. Collection Les Méthodes. Hermann, Paris. 1979. 362 p.
 
\end{thebibliography}





\end{document}